\documentclass[aps,prd,twocolumn,eqsecnum,showpacs,amsmath]{revtex4}

\usepackage{subfigure}
\usepackage[dvips]{color,graphicx}
\usepackage{amsfonts,amssymb,theorem}
\newcommand{\D}{{\rm d}}

{\theorembodyfont{\upshape}
}
{\theorembodyfont{\upshape}
}
{\theorembodyfont{\upshape}
}
{\theorembodyfont{\upshape}
}
{\theorembodyfont{\upshape}
}
{\theorembodyfont{\upshape}
}

\newcommand{\dalm}{\kern1pt\vbox{\hrule height 0.9pt\hbox{\vrule width
0.9pt\hskip 2.5pt\vbox{\vskip 5.5pt}\hskip 3pt\vrule width 0.3pt}\hrule height
0.3pt}\kern1pt}

\begin{document}
%\thispagestyle{empty}

%<<<<<<<<<<<<< TITLE >>>>>>>>>>>>>>>%
\title{
Exact dynamical AdS black holes and wormholes with a Klein-Gordon field
}

%<<<<<<<<<<<<< AUTHOR >>>>>>>>>>>>>>>%
\author{Hideki Maeda}
\email{hideki@cecs.cl}

%<<<<<<<<<<<<< ADDRESS >>>>>>>>>>>>>>>%

\address{ 
Centro de Estudios Cient\'{\i}ficos (CECs), Arturo Prat 514, Valdivia, Chile
}

%<<<<<<<<<<<<< DATE >>>>>>>>>>>>>>>%
\date{\today}

%======================================%
%<<<<<<<<<<<<< ABSTRACT >>>>>>>>>>>>>>>%
%======================================%
\begin{abstract} 
We present several classes of exact solutions in the Einstein-Klein-Gordon system with a cosmological constant.
The spacetime has spherical, plane, or hyperbolic symmetry and the higher-dimensional solutions are obtained in a closed form only in the plane symmetric case.
Among them, the class-I solution represents an asymptotically locally anti-de~Sitter (AdS) dynamical black hole or wormhole.
In four and higher dimensions, the generalized Misner-Sharp quasi-local mass blows up at AdS infinity, inferring that the spacetime is only locally AdS.
In three dimensions, the scalar field becomes trivial and the solution reduces to the Ba\~nados-Teitelboim-Zanelli black hole.
\end{abstract}

%<<<<<<<<<<<<< PACS NUMBER >>>>>>>>>>>>>>>%
% CECS-PHY-12/03
\maketitle

%======================================%
%<<<<<<<<<<<< SECTION I  >>>>>>>>>>>>>>%
%======================================%
\section{Introduction}
In comparison with stationary black holes, understanding of dynamical black holes is still far from clear.
Although there is a lot of potentially interesting subjects such as thermodynamical properties, dynamical stability, or Hawking radiation, the absence of the preferred time direction makes them intractable. 
Also, lack of concrete exact solutions in simple systems is one of the main reasons for the relatively slow progress.

In the present paper, among others, we focus on asymptotically anti-de~Sitter (AdS) dynamical black holes.
In the history of gravitation physics, AdS black holes had been considered unrealistic and eccentric configurations for a long time.
However, they stepped into the limelight by the discovery of the AdS/CFT duality~\cite{ads/cft}.
Now AdS black holes have new significance as a stage to study strongly coupled gauge theories and occupy a central position of research in string theory.

The motivation of the study in the present paper is twofold.
One comes from the AdS/CFT duality.
While a static AdS black hole corresponds to the field theory at the boundary which has finite temperature in equilibrium, a dynamical AdS black hole would correspond to some field theory in the non-equilibrium state.
Actually, an asymptotically AdS spacetime has been studied in a dynamical setting as a holographic dual to the Bjorken flow~\cite{kmno}.
While the dynamical spacetime in~\cite{kmno} was constructed perturbatively, exact dynamical AdS black holes are desirable to derive more specific results.

The second motivation comes from the recently-found dynamical instability of the AdS spacetime.
Although the AdS vacuum is known to be stable at the linear level, its nonlinear instability was numerically found with a massless Klein-Gordon field in arbitrary dimensions~\cite{br2011}. (See also~\cite{grafinkle2012}.)
It was both numerically and analytically supported that an AdS black hole forms as a result of this instability~\cite{br2011,dhs2011}.
However, there is an argument that static AdS black holes are also unstable at the nonlinear level~\cite{horowitz}. (See also~\cite{hs2011}.)
Therefore, the final fate of the instability of the AdS vacuum or a static AdS black hole is still not clear at present. 
In this context, not only a static configuration but also a dynamical configuration is the candidate of the final state.
An example is an oscillating or time-periodic spacetime~\cite{dafermos2004}.
Therefore, an exact dynamical black-hole solution might represent the final state or an intermediate stage during the time evolution and must be useful for further study.

In the present paper, we consider spacetimes with spherical, plane, or hyperbolic symmetry in arbitrary dimensions.
It is well-known in this system that the no-hair theorem holds for a wide class of scalar fields, which prohibits asymptotically flat black holes with non-trivial configuration of scalar fields~\cite{Bekenstein:1996pn}.
Here one assumes staticity to prove the no-hair theorem.
For a massless Klein-Gordon field, even a stronger result is available, namely the no-hair theorem independent of the asymptotic condition for the spacetime and the value of $\Lambda$. (See Appendix~A for the proof.)
As a result, all the known solutions with a non-trivial scalar field in this system contain naked singularities both for $\Lambda=0$~\cite{KGBHmassless1} and $\Lambda \ne 0$~\cite{KGBHmassless2}, and the only possible static black hole is the Schwarzschild(-(A)dS) black hole with a constant scalar field or its topological generalization.

Therefore, in order to obtain non-trivial black-hole solutions, one has to remove the assumption of staticity.
In four dimensions, a set of exact dynamical and inhomogeneous solutions has been obtained by many authors~\cite{lake1983,hajj1985,hp1994,vw1985,cl1987,sl1988}.
In the present paper, we generalize this set of solutions and show that some of the solutions describe a locally AdS dynamical black hole.
In the case where the Klein-Gordon field is purely imaginary, namely ghost, an AdS dynamical wormhole may be realized.

In the following section, we give our spacetime ansatz and present the solutions.
In Sec.~III, we show that the class-I solution represents an AdS dynamical black hole or wormhole.
In Sec. IV, we discuss the properties of other classes of solutions.
Concluding remarks are summarized in Sec.~V.
The scalar no-hair theorem for a massless Klein-Gordon field is shown in Appendix~A.
In Appendix~B, we present the counterpart of our solution in the case without a cosmological constant.
In Appendix~C, it is shown that the class-I solution with a real scalar field does not represent a wormhole.
Our basic notation follows~\cite{wald}.
The convention for the Riemann curvature tensor is $[\nabla _\rho ,\nabla_\sigma]V^\mu ={{\cal R}^\mu }_{\nu\rho\sigma}V^\nu$ and ${\cal R}_{\mu \nu }={{\cal R}^\rho }_{\mu \rho \nu }$.
The Minkowski metric is taken to be mostly plus sign, and Greek indices run over all spacetime indices.
We adopt the units such that only the $n$-dimensional gravitational constant $G_n$ is retained.

%======================================%
%<<<<<<<<<<<< SECTION I  >>>>>>>>>>>>>>%
%======================================%
\section{Preliminaries}
\subsection{System}
We consider the Einstein-Klein-Gordon-$\Lambda$ system in arbitrary $n(\ge 3)$ dimensions.
The field equations are $G_{\mu\nu}+\Lambda g_{\mu\nu}=\kappa_n^2T_{\mu\nu}$ and $\dalm\phi=0$, where $\kappa_n$ is defined by the $n$-dimensional gravitational constant $G_n$ as $\kappa_n:=\sqrt{8\pi G_n}$.
The energy-momentum tensor for the Klein-Gordon field is
\begin{align}
T_{\mu\nu}=\nabla_\mu\phi\nabla_\nu\phi-\frac{1}{2}g_{\mu\nu}\nabla_\rho \phi\nabla^\rho\phi.
\label{eq:stress-energy_tensor_of_scalar_field}
\end{align}

In this paper, we consider an $n$-dimensional warped product manifold $M^n \approx M^2\times K^{n-2}$ with the line element
\begin{eqnarray}
ds^2&=&g_{\mu\nu}dx^\mu dx^\nu \nonumber \\
&=&g_{AB}(y)dy^Ady^B+R(y)^2\gamma_{ij}(z)dz^idz^j,\label{sol2}
\end{eqnarray}
where $g_{AB}$ is a Lorentzian metric on $M^2$ and $R$ is a scalar on $M^2$.
$K^{n-2}$ is an $(n-2)$-dimensional unit space of constant curvature, where $k$ denotes its curvature taking the values
$1$, $0$, and $-1$, and $\gamma_{ij}(z)$ is the metric on $K^{n-2}$.
Namely the Riemann tensor on $K^{n-2}$ is given by 
\begin{eqnarray}
{}^{(n-2)}{\cal R}{}_{ijkl}=k(\gamma_{ik}\gamma_{jl}-\gamma_{il}\gamma_{jk}),
\end{eqnarray}
where the superscript $(n-2)$ means the geometrical quantity on $K^{n-2}$.

The generalized Misner-Sharp quasi-local mass is a scalar on $M^2$ defined by 
\begin{align}
\label{qlm}
m := \frac{(n-2)V_{n-2}^{(k)}}{2\kappa_n^2}R^{n-3}\biggl[-{\tilde \Lambda}R^{2}+k-(D R)^2\biggl],
\end{align}  
where ${\tilde \Lambda} := 2\Lambda /[(n-1)(n-2)]$,
$D_A $ is the covariant derivative on $M^2$ and $(DR)^2:=g^{AB}(D_AR)(D_BR)$~\cite{ms1964,nakao1995,hayward1996,mn2008}.
$V_{n-2}^{(k)}$ denotes the volume of $K^{n-2}$ if it is compact and otherwise arbitrary.
$m$ has the monotonicity and positivity properties for arbitrary (positive) $V_{n-2}^{(k)}$ and is constant in vacuum~\cite{hayward1996,mn2008}.
In the asymptotically flat or AdS case, that coefficient is fixed in such a way that it converges to the global mass such as the Arnowitt-Deser-Misner mass~\cite{ADM} or Abbott-Deser mass~\cite{AD}.

In the non-static spacetime, there is no timelike Killing vector to define a natural time-slicing.
In such a case, the Kodama vector $K^\mu  :=-\epsilon ^{\mu \nu }\nabla _\nu  R$ defines a preferred time direction in the untrapped region, where $\epsilon_{\mu \nu}=\epsilon_{AB}(\D x^A)_{\mu}(\D x^B)_{\nu}$ and $\epsilon_{AB}$ is a volume element of $(M^2, g_{AB})$~\cite{kodama1980}. 
The Kodama vector is timelike (spacelike) in the untrapped (trapped) region defined by $(DR)^2>(<)0$.
On the trapping horizon defined by $(DR)^2=0$, the Kodama vector becomes null.

\subsection{Generalized Lake solution}
In the present paper, we consider solutions in the following form:
\begin{align}
ds^2=&H(\rho)^{-2}\biggl[-dt^2+d\rho^2+S(t)\gamma_{ij}(z)dz^idz^j\biggl], \label{g-sussman} \\
H(\rho)=& \left\{
\begin{array}{ll}
\displaystyle{\sqrt{-{\tilde \Lambda}}\sin \rho} & \mbox{[class-I~($\Lambda<0$)]},\\
\displaystyle{\sqrt{-{\tilde \Lambda}}\rho} & \mbox{[class-II~($\Lambda<0$)]},\\
\displaystyle{\sqrt{-{\tilde \Lambda}}\sinh \rho} &\mbox{[class-III~($\Lambda<0$)]},\\
\displaystyle{\sqrt{{\tilde \Lambda}}\cosh \rho} &\mbox{[class-III~($\Lambda>0$)]}.
\end{array} \right. 
\end{align}
The physical domain is the region with $S>0$.
The areal radius is given by $R=(\varepsilon H)^{-1}S^{1/2}$, where $\varepsilon=\pm 1$ is chosen such that $R$ is non-negative.
These classes of solutions have been investigated as solutions with a stiff fluid, which is equivalent to a massless Klein-Gordon field if the gradient of the scalar field is timelike~\cite{madsen1988}. 
They were first obtained by Lake~\cite{lake1983} for $n=4$ and $k=1$ and independently obtained by other authors~\cite{hajj1985,hp1994,vw1985}.
The global structure and physical properties were investigated in~\cite{sussman1991}.
The solutions with $n=4$ and general $k$ were obtained by Collins and Lang~\cite{cl1987} and also in~\cite{sl1988}.
Keeping in mind this history, we call these classes of solutions the {\it generalized Lake solution} in the present paper.

The system reduces to the following master equation for $S$:
\begin{align}
4k(n-3)S+(n-4){\dot S}^2+4(n-2)wS^2+2{\ddot S}S=0,\label{master}
\end{align}
where a dot denotes the derivative with respect to $t$ and the constant $w$ is $1$, $0$, and $-1$ for class-I, -II, and -III, respectively.
The Klein-Gordon field is given as
\begin{align}
\phi=\pm \sqrt{\frac{(n-2)(n-3)}{\kappa_n^2}}\int^t \sqrt{\frac{k}{S}+\frac{{\dot S}^2}{4S^2}+w}d{\bar t}.
\end{align}  
The scalar field is homogeneous $\phi=\phi(t)$ in our coordinate system.

The energy-momentum tensor has the form of $T^\mu_{~~\nu}=\mbox{diag}(-\mu,\mu,\cdots, \mu)$, where $\mu=(1/2)H^2{\dot \phi}^2$ is the energy density of the scalar field.
Important physical quantities are given as
\begin{align}
\mu=&\frac12 H^2{\dot \phi}^2 \\
=&\frac{(n-2)(n-3)H^2}{2\kappa_n^2}\biggl(\frac{k}{S}+\frac{{\dot S}^2}{4S^2}+w\biggl),\label{mu}\\
m =&\frac{(n-2)V_{n-2}^{(k)}S^{(n-1)/2}}{2\kappa_n^2(\varepsilon H)^{n-3}}\biggl(\frac{k}{S}+\frac{{\dot S}^2}{4S^2}+w\biggl),\\
(DR)^2=&-S\biggl(\frac{{\dot S}^2}{4S^2}-\frac{{H'}^2}{H^2}\biggl),
\end{align}  
where a prime denotes the derivative with respect to $\rho$.
$\mu \ge 0$ and $m>0$ are satisfied in the spherically symmetric class-I solution ($k=w=1$).
Using $\epsilon_{t\rho}=H^{-2}$, the Kodama vector is written as
\begin{align}
K^\mu\frac{\partial}{\partial x^\mu}=&-\varepsilon H'S^{1/2}\frac{\partial}{\partial t}-\frac12 \varepsilon HS^{-1/2}{\dot S}\frac{\partial}{\partial \rho}.
\end{align}

The master equation (\ref{master}) is solved analytically in three and four dimensions for any $k$ but only for $k=0$ in higher dimensions, which will be presented later.
In order to see the qualitative property of the solution, we introduce a new variable $X:=S^{(n-2)/2}$ and write the master equation as
\begin{align}
{\ddot X}=-k(n-2)(n-3)X^{(n-4)/(n-2)}-(n-2)^2wX. 
\end{align}
This equation is integrated by parts to give
\begin{align}
E=&\frac12{\dot X}^2+V_{\rm (k)}(X), \label{master-X}\\
V_{\rm (k)}(X):=&\frac{(n-2)^2}{2}\biggl(kX^{2(n-3)/(n-2)}+wX^2\biggl),\label{V}
\end{align}
where $E$ is an integration constant.
This is a simple one-dimensional potential problem for the variable $X(t)(\ge 0)$.

Using the following expression,
\begin{align}
\frac{k}{S}+\frac{{\dot S}^2}{4S^2}+w=\frac{2E}{(n-2)^2X^{2}},
\end{align}
we obtain simple expressions of $\phi$, $\mu$ and $m$:
\begin{align}
\phi=&\pm \sqrt{\frac{2(n-3)E}{(n-2)\kappa_n^2}}\int^t \frac{d{\bar t}}{S({\bar t})^{(n-2)/2}},\\
\mu=&\frac{(n-3)EH^2}{(n-2)\kappa_n^2S^{n-2}},\label{mu2} \\
m =&\frac{E V_{n-2}^{(k)}}{(n-2)\kappa_n^2(\varepsilon H)^{n-3}S^{(n-3)/2}}.\label{m2}
\end{align}  
It is clear that the energy density of the scalar field and the quasi-local mass are positive (negative) for $E>(<)0$ and then the scalar field is real (purely imaginary, namely ghost).
In three dimensions, the scalar field becomes trivial and we have $\mu=0$ and $m=$constant.
The spacetime is then locally (A)dS.

\subsection{AdS infinity}
It is shown that, in the case of $\Lambda<0$, $H(\rho_\infty)=0$ corresponds to AdS infinity.
(In contrast, $H$ cannot be zero for $\Lambda>0$ in the class-III solution.)
Indeed, 
\begin{eqnarray}
\lim_{\rho\to \rho_\infty}R^{\mu\nu}_{~~~\rho\sigma}={\tilde\Lambda}(\delta^\mu_\rho\delta^\nu_\sigma-\delta^\mu_\sigma\delta^\nu_\rho)
\end{eqnarray}
is satisfied.
We actually show that the affine parameter $\lambda$ blows up at $\rho=\rho_\infty$ along null geodesics.
In the spacetime (\ref{g-sussman}), there is a conformal Killing vector ${\xi}_\mu dx^\mu=H^{-2}d\rho$ satisfying the conformal Killing equation:
\begin{align}
{\cal L}_\xi g_{\mu\nu}=&2\psi g_{\mu\nu},\qquad \psi:=-\frac{H'}{H}.
\end{align}
Along a null geodesic, with its tangent vector $k^\mu$, there is a conserved quantity $C_{(\xi)}:=\xi_\mu k^\nu$.
The following expression
\begin{align}
C_{(\xi)}=&H^{-2}\frac{d\rho}{d\lambda}
\end{align}
is integrated to give
\begin{align}
\label{geodesic}
\frac{1}{C_{(\xi)}(\lambda-\lambda_0)}=& \left\{
\begin{array}{ll}
\displaystyle{{\tilde \Lambda}\tan \rho} & \mbox{[class-I]},\\
\displaystyle{{\tilde \Lambda}\rho} & \mbox{[class-II]},\\
\displaystyle{{\tilde \Lambda}\tanh \rho} &\mbox{[class-III~($\Lambda<0$)]},\\
\displaystyle{{\tilde \Lambda}/\tanh \rho} &\mbox{[class-III~($\Lambda>0$)]}.
\end{array} \right. 
\end{align}
Therefore, $H=0$ for $\Lambda<0$ corresponds to $\lambda\to \infty$.
AdS infinity is given by $\rho=0$ in the class-II and -III solutions, and by $\rho=N\pi$ in the class-I solution, where $N$ is an integer.

It is seen in Eqs.~(\ref{mu2}) and (\ref{m2}) that the spacetime is indeed vacuum at AdS infinity ($H=0$), but $m$ blows up there.
The quasi-local mass $m$ with $k=1$ converges to the Abbott-Deser mass in the asymptotically AdS spacetime~\cite{mn2008} under the Henneaux-Teitelboim fall-off condition~\cite{HT1985}. (See~\cite{HIM2005} for the higher-dimensional version.) 
Its contraposition means that if $m$ blows up, the fall-off rate is slower than the Henneaux-Teitelboim condition and the spacetime is asymptotically only locally AdS~\cite{asymplocalAdS}.

\subsection{Static solution}
There is a static solution $S=S_0$ of the master equation~(\ref{master}) in the case of $kw<0$ in four and higher dimensions:
\begin{align}
S_0=&-\frac{k(n-3)}{(n-2)w},\\
\pm(\phi-\phi_0)=&\sqrt{-\frac{(n-2)w}{\kappa_n^2}}t,
\end{align}  
where $\phi_0$ is constant.
The energy density and the quasi-local mass are given by
\begin{align}
\mu=&-\frac{(n-2)wH^2}{2\kappa_n^2},\\
m =&-\frac{(n-2)wV_{n-2}^{(k)}S_0^{(n-1)/2}}{2(n-3)\kappa_n^2(\varepsilon H)^{n-3}}.
\end{align}  
While the metric is static, the scalar field is time-dependent.
In the class-III solution with $k=1$, the Klein-Gordon field is real, while it is ghost in the class-I solution with $k=-1$.

We don't present the detailed analysis for this static solution, but the Penrose diagram is Fig.~1(d) for the class-I solution with $k=-1$ and the solution represents a static AdS wormhole.
(See~\cite{ellis1973} for the wormhole solution without $\Lambda$.)
The Penrose diagram for the class-III solution with $k=1$ is Fig.~2(a) for $\Lambda<0$ and Fig.~2(f) for $\Lambda>0$.
Hereafter we don't consider the static case.

%======================================%
%<<<<<<<<<<<< SECTION I  >>>>>>>>>>>>>>%
%======================================%
\section{Class-I solution}
We are interested in the class-I solution because the coordinate system covers the maximally extended spacetime and describes an asymptotically locally AdS black hole or wormhole. 

In four dimensions, $S$ is given by 
\begin{align}
S(t)=\frac12 (-k+2C_1\sin 2t),
\end{align}
where $C_1$ is a constant.
The energy density and the quasi-local mass are given by 
\begin{align}
\mu=\frac{(4C_1^2-k^2)H^2}{4\kappa_4^2S^2},\quad m = \frac{V_{2}^{(k)}(4C_1^2-k^2)}{4\kappa_4^2\varepsilon H S^{1/2}}.
\end{align}  
The energy density is positive (negative) for $4C_1^2>(<)k^2$.
The AdS vacuum is realized for $k=1,-1$ with $4C_1^2=k^2$.
(For $k=0$, $C_1=0$ is not allowed since it gives $S\equiv 0$.)
The scalar field with positive energy density is given by
\begin{widetext}
\begin{align}
\pm (\phi-\phi_0)=& \left\{
\begin{array}{ll}
\displaystyle{\sqrt{\frac{1}{2\kappa_4^2}}\ln\biggl|\frac{\sqrt{4C_1^2-k^2}+(-k\tan t+2C_1)}{\sqrt{4C_1^2-k^2}-(-k\tan t+2C_1)}\biggl|} & \mbox{[for $k=1,-1$]},\\
\displaystyle{\sqrt{\frac{1}{2\kappa_4^2}}\ln\biggl|\frac{1-\cos 2t}{\sin 2t}\biggl|} & \mbox{[for $k=0$]}.
\end{array} \right. 
\end{align}
\end{widetext}
The scalar field with negative energy density is given as  
\begin{align}
\pm (\phi-\phi_0)=i\sqrt{\frac{2}{\kappa_4^2}}\arctan\biggl(\frac{-k\tan t+2C_1}{\sqrt{k^2-4C_1^2}}\biggl),
\end{align}
where $i^2=-1$.

In arbitrary dimensions, $S$ and $\phi$ for the class-I solution are given in closed forms only for $k=0$:
\begin{align}
S(t)=&C_1[\sin(n-2)t]^{2/(n-2)},\\
\pm(\phi-\phi_0)=&\sqrt{\frac{n-3}{(n-2)\kappa_n^2}}\ln\biggl|\frac{1-\cos(n-2)t}{\sin(n-2)t}\biggl|.
\end{align}
The energy density and the quasi-local mass are given by 
\begin{align}
\mu=&\frac{(n-2)(n-3)H^2}{2\kappa_n^2[\sin(n-2)t]^2},\\
m =&\frac{(n-2)V_{n-2}^{(0)}C_1^{(n-1)/2}}{2\kappa_n^2(\varepsilon H)^{n-3}[\sin(n-2)t]^{(n-3)/(n-2)}}.
\end{align}  
In three dimensions ($n=3$), we obtain $\mu\equiv 0$ and $m=$constant and the solution represents a Ba\~nados-Teitelboim-Zanelli black hole in the non-standard coordinates~\cite{BTZ}.

\begin{figure}[htbp]
\begin{center}
%\rotatebox{-90}{
\includegraphics[width=1.0\linewidth]{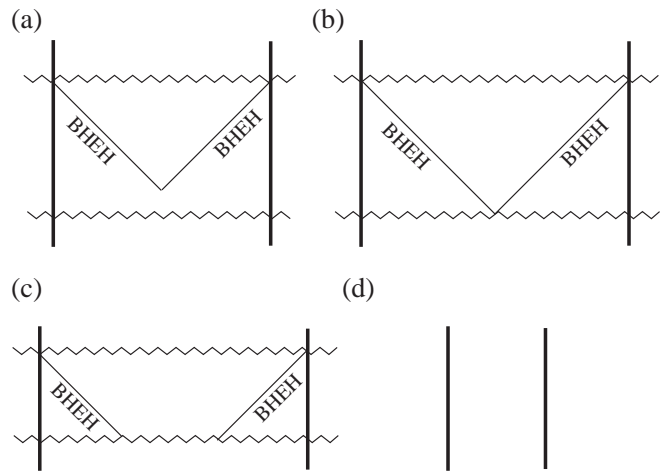}
%\subfigure[]{\includegraphics[width=0.6\linewidth]{Roberts-lambda(C)1.eps}}
%\subfigure[]{\includegraphics[width=0.6\linewidth]{Roberts-lambda(C)2.eps}}
%\subfigure[]{\includegraphics[width=0.6\linewidth]{Roberts-lambda(C)3.eps}}
%}
\caption{\label{AdSBH}
Portions of the $(\rho,t)$ plane covering the maximally extended spacetime of the class-I solution.
The corresponding Penrose diagrams have the same structures.
A zigzag and a thick line correspond to a curvature singularity and AdS infinity, respectively.
BHEH stands for the black-hole event horizon.
Figures.~(a)--(c) represent a black hole, while Fig.~(d) represents a wormhole.
}
\end{center}
\end{figure}

It is not difficult to understand the causal structure of the spacetime (\ref{g-sussman}).
$S(t)=0$ corresponds to curvature singularity, of which existence depends on the parameters.
Since the metric on $(M^2,g_{AB})$ in the solution (\ref{g-sussman}) is conformally flat, a light ray runs along a 45-degree straight line in the $(\rho,t)$-plane.
The Penrose diagrams for this solution are presented in Fig.~\ref{AdSBH}. (See Table~\ref{table:AdS}.)
The spacetime represents a dynamical black hole or wormhole depending on the parameters.
%------------<fig>---------------------------
%--------------<fig>-----------------------
\begin{center}
%-------------- TABLE ---------------%
\begin{table}[h]
\caption{\label{table:AdS} The Penrose diagrams for the class-I solution with positive energy density and $n\ge 4$. 
In the case of $k=- 1$ with negative energy density, the Penrose diagram is Fig.~1(d).}
\begin{tabular}{l@{\qquad}c@{\qquad}c}
\hline \hline
  & $n=4$ & $n\ge 5$    \\\hline
$k=0$ & Fig.~1(b) & Fig.~1(c)  \\ \hline
$k=1$ & Fig.~1(c) & Figs.~1(a), (b), or (c)  \\ \hline
$k=-1$ & Fig.~1(a) & Figs.~1(a), (b), or (c)  \\ 
\hline \hline
\end{tabular}
\end{table} 
\end{center}
%------------------------------------%

\subsection{Dynamical AdS black holes}
If the energy density is positive, there are spacelike curvature singularities given by $S(t)=0$. 
As a result, the solution represents a dynamical black hole. 
Since both $H$ and $S$ are periodic, the $(\rho,t)$ plane is divided by singularities and AdS infinities into an infinite number of portions.
The physical portions with positive $S$ are all equivalent.

First let us see the case with $k=0$.
Without loss of generality, we assume $C_1>0$ and consider a physical portion defined by $t=(0,\pi/(n-2))$ and $\rho=(0,\pi)$, which covers the maximally extended spacetime.
The event horizon in this portion is given by $t=\rho-(n-3)/(n-2)\pi$ and $t=-\rho+\pi/(n-2)$ and the Penrose diagram is (b) in Fig.~\ref{AdSBH} for $n=4$ and (c) for $n\ge 5$.

On the other hand, in the case with $k=1,-1$ in four dimensions, the period of $t$ in a physical portion is different.
The period is shorter (longer) than $\pi/2$ for $k=1~(k=-1)$.
Hence, the Penrose diagram is (c) in Fig.~\ref{AdSBH} for $k=1$ and (a) for $k=-1$.

For the class-I solution, the trapping horizon is determined by
\begin{align}
(DR)^2=&-S\biggl(\frac{{\dot S}^2}{4S^2}-\frac{1}{\tan^2\rho}\biggl)=0.
\end{align}  
In the case of $k=0$, the trapping horizon is expressed in a simple form as
\begin{align}
\tan(n-2)t=\pm \tan \rho.
\end{align}  
Thus, the trapping horizon is also drawn by spacelike straight lines in the $(\rho,t)$ plane.
For $n=4$, the trapping horizon is given by 
\begin{align}
\frac{-k+2C_1\sin 2t}{2C_1\cos 2t}=\pm \tan\rho.
\end{align}  
In order to show its signature, we study the induced metric on the trapping horizon in the $(\rho,t)$ plane:
\begin{align}
ds^2=\frac{(4C_1^2-k^2)(k^2+12C_1^2-8C_1k\sin 2t)}{H(\rho)^2(k^2+4C_1^2-4C_1k\sin 2t)^2}dt^2.
\end{align}
A portion of the trapping horizon with $ds^2>(<)0$ is spacelike (timelike).
We can show that the trapping horizon is spacelike if $4C_1^2-k^2>0$, namely the energy density of the scalar field is positive.
It is obvious for $k=0,-1$ by the following expression:
\begin{align}
k^2+12C_1^2-8C_1k\sin 2t=3(4C_1^2-k^2)-8kS. \label{key-t}
\end{align}
For $k=1$, using the inequality $0<S\le (-1+2|C_1|)/2$, the above equation is evaluated as
\begin{align}
12|C_1|^2-8|C_1|+1 \le 3(4C_1^2-1)-8S<3(4C_1^2-1).
\end{align}
Since the lower bound is positive for $4C_1^2-1>0$, it is concluded that the trapping horizon is spacelike.

In the case with $k=\pm1$ and $n\ge 5$, the solution is not obtained in a closed form, but we can prove that it represents an AdS black hole if the energy density of the scalar field is positive, namely $E>0$.
For $k=1$, the potential (\ref{V}) is monotonically increasing for $X\ge 0$ and hence the solution exists only for $E>0$.
Then, the domain of $t$ in the maximally extended spacetime of the solution is given by $t_0 <t< t_0+ T$, where $X(t_0)=X(t_0+ T)=0$.
This is also the case for $k=-1$ with $E\ge 0$.
The period $T$ is defined by
\begin{align}
T:=&2\int_{X=0}^{X=X_{\rm b(k)}}\frac{dX}{\sqrt{2(E-V_{\rm (k)}(X))}},
\end{align}
where $X_{\rm b(k)}$ is defined by $E=V_{\rm (k)}(X_{\rm b(k)})$.
The spacetime admits a wormhole throat if $T\ge \pi$ because the period of the coordinate $\rho$ is $\pi$, however it is not allowed if the scalar field has positive energy density. (See Appendix~C for the proof.)
Since $t=t_0$ and $t=t_0+T$ are both spacelike curvature singularities, the corresponding Penrose diagram is Figs.~1(a), 1(b), and 1(c) for $\pi/2<T<\pi$, $T=\pi/2$, and $0<T<\pi/2$, respectively.
Although the diagrams are different depending on the value of $T$, the solution represents a dynamical AdS black hole.

\subsection{Dynamical AdS wormholes}
In the case of $k=-1$ in four dimensions, if $4C_1^2<k^2$, then the energy density is negative and $S$ is positive definite for $-\infty<t<\infty$.
(There is no physical solution for $k=1$ because $S$ is negative definite if $4C_1^2<k^2$.)
The Klein-Gordon field then becomes ghost and there is no curvature singularity in the spacetime.
As a result, the spacetime is a dynamical AdS wormhole described by the Penrose diagram (d) in Fig.~\ref{AdSBH}.

Let us discuss the signature of the trapping horizon.
Since $(1-2|C_1|)/2\le S\le (1+2|C_1|)/2$ is satisfied in the present case, the right-hand side of Eq.~(\ref{key-t}) is evaluated as
\begin{align}
12|C_1|^2-8|C_1|+1 \le 3(4C_1^2-1)+8S \le 12|C_1|^2+8|C_1|+1.\label{ineq1}
\end{align}
While the upper bound is positive definite, the lower bound is positive for $0\le |C_1|<1/6$ and negative for $1/6< |C_1|<1/2$.
Therefore, the trapping horizon is timelike for $0\le |C_1|<1/6$.
In contrast, the trapping horizon has both timelike and spacelike portions for $1/6< |C_1|<1/2$.
For $|C_1|=1/6$, the lower bound of Eq.~(\ref{ineq1}) is $0$ and the trapping horizon becomes null only at the moment of the bounce characterized by ${\dot S}=0$ and ${\ddot S}>0$.

It is shown that an AdS wormhole is realized also for $k=-1$ and $n\ge 5$ if $E<0$; namely the energy density is negative.
In the case of $k=-1$, the potential (\ref{V}) in the master equation has a negative minimum $V_{\rm (-1)}=V_{\rm ex}(<0)$, where 
\begin{align}
V_{\rm ex}:=&-\frac{n-2}{2}\biggl(\frac{n-3}{n-2}\biggl)^{n-3}.
\end{align}  
As a result, in the solution with $E$ satisfying $V_{\rm ex}<E<0$, the value of $X$ (and hence $S$) oscillates and never becomes $0$.
Hence, the corresponding Penrose diagram is Fig.~1(d) and the spacetime describes a dynamical AdS wormhole.

%======================================%
%<<<<<<<<<<<< SECTION I  >>>>>>>>>>>>>>%
%======================================%
\section{Class-II and -III solutions}
Next let us consider the class-II and -III solutions.
We only consider the case in which the solution is obtained in a closed form.

In four dimensions, $S(t)$ is given by 
\begin{align}
S(t)=& \left\{
\begin{array}{ll}
\displaystyle{-kt^2+2D_1t+D_2} & \mbox{[class-II]},\\
\displaystyle{\frac12 (k+2D_3\cosh 2t)} & \mbox{[class-III(a)]},\\
\displaystyle{\frac12 (k+2D_4\sinh 2t)} & \mbox{[class-III(b)]},\\
\displaystyle{\frac12 (k+D_5e^{2t})} & \mbox{[class-III(c)]},
\end{array} \right. 
\end{align}
where $D_1$--$D_5$ are constants.
The class-III solution was further classified into three.
The energy density and the quasi-local mass are written as 
\begin{align}
\mu=\frac{\mu_0H^2}{4\kappa_4^2S^2},\quad m = \frac{V_{2}^{(k)}\mu_0}{4\kappa_4^2\varepsilon HS^{1/2}},
\end{align}  
where the constant $\mu_0$ is defined by
\begin{align}
\mu_0:=&4S^{n-2}\biggl(\frac{k}{S}+\frac{{\dot S}^2}{4S^2}+w\biggl), \label{mu0}\\
=& \left\{
\begin{array}{ll}
\displaystyle{4(kD_2+D_1^2)} & \mbox{[class-II]},\\
\displaystyle{k^2-4D_3^2} & \mbox{[class-III(a)]},\\
\displaystyle{k^2+4D_4^2} & \mbox{[class-III(b)]},\\
\displaystyle{k^2} & \mbox{[class-III(c)]}.
\end{array} \right. 
\end{align}
The relation between $\mu_0$ and the energy constant $E$ in the master equation (\ref{master-X}) is $\mu_0=8E/(n-2)^2$.
The Klein-Gordon field with positive energy density is given by
\begin{widetext}
\begin{align}
\pm(\phi-\phi_0)=& \left\{
\begin{array}{ll}
\displaystyle{\sqrt{\frac{1}{2\kappa_4^2}}\ln\biggl|\frac{\sqrt{D_1^2+kD_2}+(D_1-kt)}{\sqrt{D_1^2+kD_2}-(D_1-kt)}\biggl|} & \mbox{[class-II ($k=1,-1$)]},\\
\displaystyle{\sqrt{\frac{1}{2\kappa_4^2}}\ln|2D_1t+D_2|} & \mbox{[class-II ($k=0$)]},\\
\displaystyle{\sqrt{\frac{1}{2\kappa_4^2}}\ln\biggl|\frac{\sqrt{k^2-4D_3^2}+(k-2D_3)\tanh t}{\sqrt{k^2-4D_3^2}-(k-2D_3)\tanh t}\biggl|} & \mbox{[class-III(a) ($k=1,-1$)]},\\
\displaystyle{\sqrt{\frac{1}{2\kappa_4^2}}\ln\biggl|\frac{\sqrt{k^2+4D_4^2}+(2D_4-k\tanh t)}{\sqrt{k^2+4D_4^2}-(2D_4-k\tanh t)}\biggl|} & \mbox{[class-III(b) ($k=1,-1$)]},\\
\displaystyle{\sqrt{\frac{1}{2\kappa_4^2}}\ln\biggl|\frac{1+e^{2t}}{1-e^{2t}}\biggl|} & \mbox{[class-III(b) ($k=0$)]},\\
\displaystyle{\sqrt{\frac{1}{2\kappa_4^2}}\ln|D_5+ke^{-2t}|} & \mbox{[class-III(c)]}.
\end{array} \right. 
\end{align}
The scalar field with negative energy density is given as
\begin{align}
\pm(\phi-\phi_0)=& \left\{
\begin{array}{ll}
\displaystyle{i\sqrt{\frac{2}{\kappa_4^2}}~\arctan\biggl(\frac{D_1-kt}{\sqrt{-(D_1^2+kD_2)}}\biggl)} & \mbox{[class-II ($k=1,-1$)]},\\
\displaystyle{i\sqrt{\frac{2}{\kappa_4^2}}~\arctan\biggl(\frac{(k-2D_3)\tanh t}{\sqrt{4D_3^2-k^2}}}\biggl) & \mbox{[class-III(a) ($k=1,-1$)]},\\
\displaystyle{i\sqrt{\frac{2}{\kappa_4^2}}~\arctan(e^{2t})} & \mbox{[class-III(a) ($k=0$)]},\\
\displaystyle{i\sqrt{\frac{2}{\kappa_4^2}}~\arctan\biggl(\frac{2D_4-k\tanh t}{\sqrt{-(k^2+4D_4^2)}}}\biggl) & \mbox{[class-III(b) ($k=1,-1$)]}.
\end{array} \right. 
\end{align}
\end{widetext}

In higher dimensions, the metric function $S$ of the non-vacuum solution is obtained in a closed form only for $k=0$:
\begin{align}
S(t)=& \left\{
\begin{array}{ll}
\displaystyle{[(n-2)D_1t+D_2]^{2/(n-2)}} & \mbox{[class-II]},\\
\displaystyle{D_3[\cosh(n-2)t]^{2/(n-2)}} & \mbox{[class-III(a)]},\\
\displaystyle{D_4[\sinh(n-2)t]^{2/(n-2)}} & \mbox{[class-III(b)]}.
\end{array} \right. 
\end{align}
The energy density and the quasi-local mass are written as
\begin{align}
\mu=&\frac{(n-2)(n-3)\mu_0H^2}{8\kappa_n^2S^{n-2}},\\
m =& \frac{(n-2)V_{n-2}^{(0)}\mu_0}{8\kappa_n^2(\varepsilon H)^{n-3}S^{(n-3)/2}},
\end{align}  
where $\mu_0$ defined by Eq.~(\ref{mu0}) is 
\begin{align}
\mu_0=& \left\{
\begin{array}{ll}
\displaystyle{4D_1^2} & \mbox{[class-II]},\\
\displaystyle{-4D_3^{n-2}} & \mbox{[class-III(a)]},\\
\displaystyle{4D_4^{n-2}} & \mbox{[class-III(b)]}.
\end{array} \right. 
\end{align}
The class-III(c) with $k=0$ is vacuum, so we don't treat here.
The scalar field is given as
\begin{widetext}
\begin{align}
\pm(\phi-\phi_0)=& \left\{
\begin{array}{ll}
\displaystyle{\sqrt{\frac{n-3}{(n-2)\kappa_n^2}}\ln|(n-2)D_1t+D_2|} & \mbox{[class-II]},\\
\displaystyle{i\sqrt{\frac{4(n-3)}{(n-2)\kappa_n^2}}~\mbox{arctan}(e^{(n-2)t})} & \mbox{[class-III (a)]},\\
\displaystyle{\sqrt{\frac{n-3}{(n-2)\kappa_n^2}}\ln\biggl|\frac{1+e^{(n-2)t}}{1-e^{(n-2)t}}\biggl|} & \mbox{[class-III (b)]}.
\end{array} \right. 
\end{align}
\end{widetext}
Since the energy density is negative, the scalar field is ghost in the class-III(a) solution.

As the class-I solution, the $(\rho,t)$ plane for the class-II or -III solution is divided by the lines of curvature singularities or AdS infinity but into a finite number of portions, unlike the class-I solution.
The structure of the $(\rho,t)$ plane is classified by the number of spacelike curvature singularities given by $S(t)=0$.
(See Table~\ref{table:qlm}.)
\begin{widetext}
\begin{center}
%-------------- TABLE ---------------%
\begin{table}[h]
\caption{\label{table:qlm} The number of spacelike curvature singularities in the Penrose diagram of the $(\rho,t)$ plane for the class-II and -III solutions with $k=1,-1$ in the four-dimensional non-vacuum case.
For $k=0$ in $n (\ge 4)$ dimensions, there is one singularity in the class-II and -III(b) solutions while there is no singularity in the class-III(a) solution.}

\begin{tabular}{l@{\qquad}c@{\qquad}c@{\qquad}c}
\hline \hline
No. of singularities  & 0 & 1 & 2   \\\hline
Class-II & $D_1^2+kD_2<0$ & $D_1^2+kD_2=0$ & $D_1^2+kD_2>0$ \\ \hline
Class-III(a) & $D_3(k+2D_3)>0$ or $D_3=0$ & $k+2D_3=0$ & $D_3(k+2D_3)<0$ \\ \hline
Class-III(b) & $D_4 =0$ & $D_4 \ne 0$ & Not applicable \\ \hline
Class-III(c) & $D_5(k+D_5) > 0$ or $D_5=0$  & $D_5(k+D_5) \le 0$ with $D_5\ne 0$  & Not applicable \\ 
\hline \hline
\end{tabular}
\end{table} 
\end{center}
%------------------------------------%
\end{widetext}

The corresponding Penrose diagrams are drawn in Fig.~2.
In those diagrams, the regular coordinate boundary $\rho=\pm\infty$, which consists of null hypersurfaces in Fig.~2 drawn by dashed lines, is extendable.
This is obvious by Eq.~(\ref{geodesic}) since $\rho=\pm\infty$ corresponds to a finite value of the affine parameter $\lambda$.
Furthermore, $\mu \to 0$ and $m\to 0$ are satisfied for $\rho=\pm\infty$ along null geodesics.
This fact indicates that a variety of $C^1$ extension is possible beyond this coordinate boundary without introducing any matter field on the junction surface.
One possible extension is to attach an exact (A)dS spacetime~\cite{sussman1991}.
 
Although $\rho=\pm\infty$ is regular and extendable along null geodesics, it is singular along spacelike curves with $t=$constant, where $\mu$ blows up.
Therefore, although the regions (ii) in Fig.~2(e) and (j) are maximally extended, there is no regular Cauchy surface.
The regions (ii) in Figs.~2(b) and 2(e) and (iii) in Fig.~2(e) contain a black-hole event horizon.
%------------<fig>---------------------------
\begin{figure}[htbp]
\begin{center}
%\rotatebox{-90}{
\includegraphics[width=1.0\linewidth]{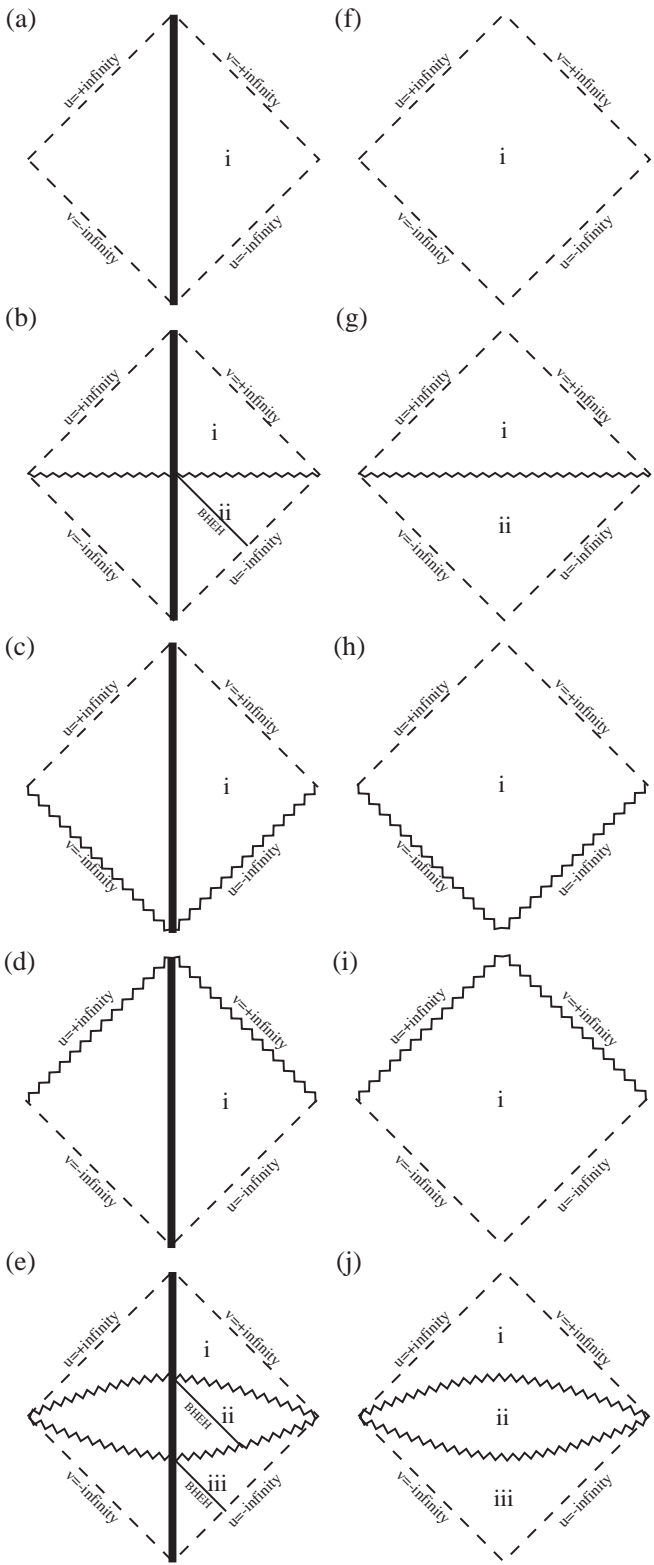}
%\subfigure[]{\includegraphics[width=0.6\linewidth]{Roberts-lambda(C)1.eps}}
%\subfigure[]{\includegraphics[width=0.6\linewidth]{Roberts-lambda(C)2.eps}}
%\subfigure[]{\includegraphics[width=0.6\linewidth]{Roberts-lambda(C)3.eps}}
%}
\caption{\label{Penrose5}
The Penrose diagrams for the class-II and -III solutions.
Figures (a)--(e) [(f)--(j)] correspond to the case with $\Lambda<(>)0$.
Figures (c), (d), (h), and (i) correspond to the special case of the class-III(c) solution with $k+D_5=0$.
Coordinate boundaries represented by dashed lines are extendable.
The $(\rho,t)$ plane is divided by singularities and AdS infinity into several portions and each of them represents a distinct spacetime.
The parameters are required to give $S>0$ for the physical spacetime and the left halves of Figs.~(a)--(e) are equivalent to the right halves. 
The region (ii) in Fig.~(e) is maximally extended and contains an event horizon, but there is no regular Cauchy surface.
}
\end{center}
\end{figure}
%--------------<fig>-----------------------

%======================================%
%<<<<<<<<<<<< SECTION I  >>>>>>>>>>>>>>%
%======================================%
\section{Summary}
We have presented a set of exact solutions in the Einstein-Klein-Gordon system with a cosmological constant in arbitrary dimensions.
The spacetime has spherical, plane, or hyperbolic symmetry and admits a spatially conformal Killing vector.
The solution is obtained in a closed form in three and four dimensions for any $k$ but only for $k=0$ in higher dimensions.
Even in the case without the explicit form, it is able to understand the qualitative properties of the solutions by analyzing the equivalent one-dimensional potential problem.
In three dimensions, the solution reduces to the locally (A)dS vacuum.

For $\Lambda<0$, the spacetime is asymptotically locally AdS.
The quasi-local mass blows up at AdS infinity while the energy density converges to zero, which infers the slow fall-off to the AdS infinity. 
Some of the solutions admit a black-hole event horizon.

In the class-I solution, the coordinate system covers the maximally extended spacetime and the solution with a real scalar field describes a dynamical AdS black hole.
If the scalar field is ghost, the solution represents a dynamical AdS wormhole.
While the solution with $k=-1$ in four dimensions represents the dynamical formation of a black hole, the black hole is eternal in the case of $k=1$ in four dimensions and $k=0$ in $n(\ge 4)$ dimensions.
It is still not clear whether the black hole is eternal or not in the case of $k=\pm 1$ in higher dimensions.

For the class-II and -III solutions, we have analyzed the global structures in four dimensions and in higher dimensions with $k=0$.
The regular coordinate boundary is extendable and the $C^1$ extension beyond it would be possible.
There are several cases where the spacetime contains a black-hole event horizon, however, the coordinate system does not cover the maximally extended spacetime or there is no regular Cauchy surface in the spacetime.

In summary, the class-I solution may be a good model of a dynamical AdS black hole for further investigations.
Thermodynamical properties, dynamical stability, and the Hawking radiation are interesting subjects to study, of which results will shed light on dynamical properties of AdS black holes.

\subsection*{Acknowledgements}
The author thanks Julio Oliva, Takashi Torii, Kei-ichi Maeda, Cristi{\'a}n Mart\'{\i}nez, and especially Fabrizio Canfora for useful comments and discussions. 
This work has been funded by the Fondecyt grants 1100328, 1100755 (HM) and by the Conicyt grant "Southern Theoretical Physics Laboratory" ACT-91. 
This work was also partly supported by the JSPS Grant-in-Aid for Scientific Research (A) (22244030).
The Centro de Estudios Cient\'{\i}ficos (CECs) is funded by the Chilean Government through the Centers of Excellence Base Financing Program of Conicyt.

\appendix

\section{No-hair theorem for a Klein-Gordon field}
In this appendix, we present a simple proof of the no-hair theorem for a Klein-Gordon field; there is no Killing horizon in the spacetime represented by the metric (\ref{sol2}) if the spacetime is static and the Klein-Gordon field is static and inhomogeneous.
We note that this result is independent of the value of $\Lambda$ and the asymptotic condition for the spacetime.

We adopt the following coordinates for the static spacetime:
\begin{align}
ds^2=&-f(r)e^{-2\delta(r)}dt^2+f(r)^{-1}dr^2+R(r)^2\gamma_{ij}dz^idz^j.
\end{align}
We can replace $R$ by $r$ without loss of generality if $R$ is not constant.
For $\phi=\phi(r)$, the Klein-Gordon equation $\dalm \phi=0$ gives
\begin{align}
\frac{d}{dr}\biggl(e^{-\delta}R^{n-2}f\phi'\biggl)=0~\to~\phi'=\frac{C_\phi e^{\delta}}{fR^{n-2}},\label{KG1}
\end{align}
where $C_\phi$ is a constant and a prime denotes here the derivative with respect to $r$.
If $C_\phi$ is zero, $\phi$ is constant.
The trace of the Einstein equation gives the expression of the Ricci scalar ${\cal R}$:
\begin{align}
{\cal R}=&\kappa_n^2(\nabla\phi)^2+\frac{2n}{n-2}\Lambda \nonumber \\
=&\kappa_n^2\frac{C_\phi^2 e^{2\delta}}{fR^{2(n-2)}}+\frac{2n}{n-2}\Lambda, \label{key}
\end{align}
where we used Eq.~(\ref{KG1}).
A Killing horizon is defined by $f(r_{\rm h})=0$ with $|\delta(r_{\rm h})|<\infty$, where $r=r_{\rm h}$ is its location satisfying $0<R(r_{\rm h})<\infty$.
Equation~(\ref{key}) shows ${\cal R}\to \infty$ for $r\to r_{\rm h}$, namely a curvature singularity at $r=r_{\rm h}$, unless $C_\phi=0$.
Therefore, the Killing horizon is allowed only in the case where $\phi$ is constant.

\section{Generalized Roberts solution}
In this appendix, we present the counterpart of the generalized Lake solution in the case without a cosmological constant.
The conformal self-similarity naturally reduces to the homothetic self-similarity in the absence of a characteristic scale introduced by the cosmological constant.

First let us derive the master equation for the system.
Introducing the double-null coordinates $(u,v)$ on $(M^2, g_{AB})$ and assuming that $(M^2, g_{AB})$ is flat and $\phi=\phi({\bar z})$, where ${\bar z}:=v/u$, the line element for the homothetic self-similar spacetime is given by
\begin{align}
ds^2=&-2du({\bar z}du+ud{\bar z})+u^2{\cal S}({\bar z})\gamma_{ij}dz^idz^j.
\end{align}
The Einstein tensor and the energy-momentum tensor for the scalar field are written as
\begin{align}
G^{u}_{~u}=&\frac{n-2}{4u^2{\cal S}^2}\biggl[(n-5){\bar z}{{\cal S}'}^2-2(n-3){\cal S}{\cal S}' \nonumber \\
&-2k(n-3){\cal S}+4{\bar z}{\cal S}{\cal S}''\biggl],\\
G^{u}_{~{\bar z}}=&-\frac{n-2}{4u{\cal S}^2}({{\cal S}'}^2-2{\cal S}{\cal S}''),\quad G^{{\bar z}}_{~u}=0,\\
G^{{\bar z}}_{~{\bar z}}=&\frac{(n-2)(n-3)}{4u^2{\cal S}^2}(-2k{\cal S}-2{\cal S}{\cal S}'+{\bar z}{{\cal S}'}^2),\\
G^{i}_{~j}=&\frac{n-3}{4u^2{\cal S}^2}\biggl[(n-6){\bar z}{{\cal S}'}^2-2(n-4){\cal S}{\cal S}' \nonumber \\
&-2k(n-4){\cal S}+4{\bar z}{\cal S}{\cal S}''\biggl]\delta^i_{~j}
\end{align}
and 
\begin{align}
T^{u}_{~u}=&-u^{-2}{\bar z}{\phi'}^2,\quad T^{u}_{~{\bar z}}=-u^{-1}{\phi'}^2,\\
T^{{\bar z}}_{~u}=&0,\quad  T^{{\bar z}}_{~{\bar z}}=u^{-2}{\bar z}{\phi'}^2, \quad T^{i}_{~j}=-u^{-2}{\bar z}{\phi'}^2\delta^i_{~j},
\end{align}
where a prime denotes here the derivative with respect to ${\bar z}$.

Then, the Einstein equation gives the following master equation for ${\cal S}$:
\begin{align}
0=(n-4){\bar z}{{\cal S}'}^2-2(n-3){\cal S}{\cal S}'-2k(n-3){\cal S}+2{\bar z}{\cal S}{\cal S}''.
\end{align}
The solution for this master equation is obtained in a closed form for $n=4$ or $k=0$. 
(The solution for $n=3$ is vacuum and hence locally flat, as shown below.)
In the case of $n=4$, $\cal S$ is given by
\begin{align}
{\cal S}({\bar z})=&-k{\bar z}+E_1{\bar z}^2+E_2,
\end{align}
where $E_1$ and $E_2$ are constants.
For $k^2-4E_1E_2>0$, $\phi$ is real and given as
\begin{widetext}
\begin{align}
\pm(\phi-\phi_0)=& \left\{
\begin{array}{ll}
\displaystyle{\sqrt{\frac{1}{2\kappa_4^2}}\ln\biggl|\frac{\sqrt{k^2-4E_1E_2}+(k-2E_1{\bar z})}{\sqrt{k^2-4E_1E_2}-(k-2E_1{\bar z})}\biggl|} & \mbox{for $E_1\ne0$},\\
\displaystyle{\sqrt{\frac{1}{2\kappa_4^2}}\ln|E_2-k{\bar z}|} & \mbox{for $E_1=0$}.
\end{array} \right. 
\end{align}
\end{widetext}
For $k^2-4E_1E_2<0$, $\phi$ is ghost and given by
\begin{align}
\pm(\phi-\phi_0)=i\sqrt{\frac{2}{\kappa_4^2}}\arctan\biggl(\frac{k-2E_1{\bar z}}{\sqrt{4E_1E_2-k^2}}\biggl).
\end{align}
In the case of $k=0$, the solution is obtained as 
\begin{align}
{\cal S}=&(E_1{\bar z}^{n-2}+E_2)^{2/(n-2)},\\
\pm(\phi-\phi_0)=&\sqrt{-\frac{(n-2)(n-3)E_1E_2}{\kappa_n^2}}\int^{\bar z}\frac{{\hat z}^{(n-4)/2}}{E_1{\hat z}^{n-2}+E_2}d{\hat z}.
\end{align}
It is noted that $k^2=4E_1E_2$ gives the Minkowski spacetime.

In the double-null coordinates, the metric and the generalized Misner-Sharp mass are written as
\begin{align}
ds^2=&-2dudv+(-kuv+E_1v^2+E_2u^2)\gamma_{ij}dz^idz^j,\\
m=&\frac{-V_2^{(k)}uv(k^2-4E_1E_2)}{2\kappa_4^2\sqrt{-kuv+E_1v^2+E_2u^2}}
\end{align}
for $n=4$ and 
\begin{align}
ds^2=&-2dudv+(E_1v^{n-2}+E_2u^{n-2})^{2/(n-2)}\gamma_{ij}dz^idz^j,\\
m=&\frac{(n-2)V_{n-2}^{(0)}E_1E_2(uv)^{n-3}}{\kappa_n^2(E_1v^{n-2}+E_2u^{n-2})^{(n-3)/(n-2)}}
\end{align}
for $k=0$.
This solution with $k=1$ and $n=4$ is the Roberts solution~\cite{roberts1989}. (See also~\cite{gb1967,ont1994,brady1994,burko1997,hayward2000,ch2001,maeda2009}.)
The case with $k=1$ in arbitrary dimensions was first discussed in~\cite{frolov1999}.
This spacetime admits a homothetic Killing vector $\xi^\mu(\partial/\partial x^\mu)=u(\partial/\partial u) +v(\partial/\partial v)$ satisfying ${\cal L}_{\xi}g_{\mu\nu}=2g_{\mu\nu}$.
Since $(M^2, g_{AB})$ is flat, $u\to \pm\infty$ with constant $v$ or $v\to \pm\infty$ with constant $u$ corresponds to null infinity.
Different from the generalized Lake solution, the gradient of the scalar field may be spacelike in some domain of spacetime, where the solution is not equivalent to a solution with a stiff fluid.

\section{No wormhole in the class-I solution with real scalar field}
In this appendix, we show that the class-I solution with positive energy density does not represent a wormhole for $n\ge 4$.
This result restricts the possible Penrose diagram for the solution without the explicit form. 
Here we define a wormhole spacetime by the existence of causal curves connecting one infinity to another and prove the non-existence of such curves.
Because the orbits of timelike curves or non-radial null curves run inside the light cone at a given spacetime point in the $(\rho,t)$ plane, to show the non-existence for radial null curves is sufficient.

First we show that the areal radius $R:=S^{1/2}/(\varepsilon H)$ blows up at AdS infinity in the class-I solution for $n \ge 4$, which is characterized by ${H}(\rho)=0$ and $0 \le S(t)<\infty$.
It is obvious that $R \to \infty$ holds if $S>0$ there.
AdS infinity where $S=0$ is satisfied is more subtle but it is also the case, as shown below.

In the class-I solution with positive energy density ($E>0$), the master equation (\ref{master-X}) shows us the behavior of $X$ near $X=0$ (and hence $S=0$):
\begin{align}
S=X^{2/(n-2)} \simeq (2E)^{1/(n-2)}|t-t_0|^{2/(n-2)},
\end{align}
where $t_0$ is the time when $S=0$.
On the other hand, the metric function $H(\rho)$ behaves as $H\simeq \sqrt{-{\tilde\Lambda}}|\rho|$ near AdS infinity.
Because $d\rho=\pm dt$ is satisfied along a radial null curve, $|\rho|=|t-t_0|$ is satisfied along such a light ray going to or coming from AdS infinity with $S=0$.
Along this curve, the areal radius $R$ behaves near AdS infinity as
\begin{align}
\lim_{S,H\to 0}R=&\lim_{S,H\to 0}\frac{S^{1/2}}{\varepsilon H}\simeq \frac{ (2E)^{1/[2(n-2)]}}{\sqrt{-{\tilde\Lambda}}|t-t_0|^{(n-3)/(n-2)}}.
\end{align}
Therefore, the areal radius $R$ blows up at AdS infinity.

Now we show that there is no radial null curve connecting two distinct AdS infinity in the spacetime of the class-I solution with positive energy density.
Let us consider the Einstein equation $G_{\mu\nu}+\Lambda g_{\mu\nu}=\kappa_n^2T_{\mu\nu}$ in the double null coordinates:
\begin{align}
\D s^2 = -2e^{-f(u,v)}\D u\D v+R(u,v)^2 \gamma_{ij}\D z^i\D z^j.\label{metric-double}
\end{align}  
The $(u,u)$ and $(v,v)$ components of the Einstein equation are written as
\begin{align}
\frac{\partial_u\partial_u R}{R}+\partial_uf\frac{\partial_uR}{R}=&-\frac{\kappa_n^2}{n-2}  T_{uu}, \label{nullbasic1} \\
\frac{\partial_v\partial_v R}{R}+\partial_vf\frac{\partial_vR}{R}=&-\frac{\kappa_n^2}{n-2}  T_{vv}. \label{nullbasic2} 
\end{align}  
The null energy condition requires $T_{uu}\ge 0$ and $T_{vv} \ge 0$.

The generalized Lake solution is written in the double-null coordinates by introducing $u$ and $v$ such that
\begin{eqnarray}
\rho(u,v)=\frac{v-u}{\sqrt{2}},\qquad t(u,v)=\frac{v+u}{\sqrt{2}}.
\end{eqnarray}
The correspondence between (\ref{g-sussman}) and (\ref{metric-double}) is
\begin{align}
e^{-f(u,v)}=H(\rho(u,v))^{-2},\quad R=\frac{{S(t(u,v))}^{1/2}}{\varepsilon {H}(\rho(u,v))}.
\end{align}

If the solution represents an AdS wormhole, there is a radial null ray which travels from one AdS infinity (where $R\to \infty$) to the other.
Obviously there is at least one positive local minimum of $R$ along its orbit, which locally defines a wormhole throat~\cite{hv1998}.
Without loss of generality, this orbit is expressed by $u=u_0$ (and $z^i=$constant), where $u_0$ is a constant and the throat condition is then given by $\partial_v R=0$ with $\partial_v\partial_v R>0$.
Then, Eq.~(\ref{nullbasic2}) shows $T_{vv}<0$, the violation of the null energy condition at the throat.
The contraposition of this result shows that the class-I solution with positive energy density does not represent an AdS wormhole.

\end{document}